\documentclass[pra,amssymb,amsmath,superscriptaddress,twocolumn,longbibliography,nofootinbib]{revtex4-1}

\usepackage{color,braket,tikz,calligra}
\usepackage[T1]{fontenc}
\usetikzlibrary{arrows}

\newcommand{\blu}{\color{blue}}

\newcommand{\bla}{\color{black}}

\newcommand{\be}{\begin{equation}}
\newcommand{\ee}{\end{equation}}

\begin{document}

\title{Quantum bit  commitment and the  reality of the  quantum state}
\author{R.            Srikanth}          \email{srik@poornaprajna.org}
\affiliation{Poornaprajna Institute of  Scientific Research, Bengaluru
  560 080, India}

\begin{abstract}
Quantum   bit   commitment  (QBC)   is   insecure   in  the   standard
non-relativistic quantum cryptographic  framework, essentially because
Alice can  exploit quantum  steering to  defer making  her commitment.
Two  assumptions in  this  framework  are that:  (a)  Alice knows  the
ensembles of evidence $E$ corresponding  to either commitment; and (b)
system  $E$ is  quantum  rather  than classical.   Here,  we show  how
relaxing  assumption (a)  or  (b) can  render  her malicious  steering
operation  indeterminable or  inexistent,  respectively.  Finally,  we
present a secure  protocol that relaxes both assumptions  in a quantum
teleportation setting.  Without appeal to an ontological framework, we
argue that the protocol's security  entails the reality of the quantum
state, provided retrocausality is excluded.
\end{abstract}


\maketitle

\section{Introduction\label{sec:intro}}
Bit commitment  (BC) is a  cryptographic mistrustful task  between two
adversarial parties Alice and Bob, wherein  Alice commits a bit $a$ by
submitting  as  evidence a  quantum  system,  possibly after  multiple
rounds of communication between them,  and later she unveils $a$.  The
security  requirement is  that the  evidence must  be binding  on her,
while  hiding $a$  from  Bob  until her  unveiling.   BC is  important
because it can serve as  a primitive for important crypto-tasks, among
them   coin   flipping,   oblivious   transfer,   secure   multi-party
computation, signature  schemes and zero-knowledge proofs.   Except by
invoking  computational   assumptions,  a  trusted  third   party,  or
relativistic constraints on  signaling \cite{Ken99,Ken12,KTH+0, LKB+0,
  LKB+15,  VMH+16}, secure  BC is  conventionally not  believed to  be
possible.

We briefly recapitulate a version  of the standard insecurity argument
against (nonrelativistic)  quantum BC  (QBC) \cite{May97,LC97,cDP+13}.
Suppose  $\mathcal{E}^\alpha \equiv  \{|\tilde{\chi}_j^\alpha\rangle =
\sqrt{p_j^\alpha}  \ket{\chi^\alpha_j}\}$  denotes   the  ensemble  of
un-normalized  states  of  the   evidence,  corresponding  to  Alice's
commitment  to bit  $\alpha$, which  Alice  submits to  Bob.  For  her
commitment to be concealed from Bob, it is required that
\begin{equation}
\rho^0_B  = \rho^1_B,
\label{eq:conceal}
\end{equation}
where    $\rho^\alpha_B    =   \sum_j    |\tilde{\chi}_j^\alpha\rangle
\langle\tilde{\chi}_j^\alpha|$  and   $\alpha\in\{0,1\}$.   To  cheat,
Alice submits  the second  register of the  purification 
\begin{equation}
|\Psi\rangle \equiv \sum_j  |\phi_j^0\rangle |\tilde{\chi}_j^0\rangle
\label{eq:mlc}
\end{equation}
as her  evidence system, where  $|\phi^0_j\rangle$ are elements  of an
orthonormal  basis.   To unveil  $\alpha=0$,  she  measures the  first
register  in the  basis $\{|\phi_j^0\rangle\}$  at unveiling  time and
announces  the   outcome  $j$.   To  unveil   $\alpha=1$,  she  steers
\cite{WJD07} $E$ into the ensemble $\mathcal{E}^1$ before measurement,
i.e., she measures the first register in the basis $\{|\phi^1_j\rangle
= \mathcal{U} |\phi^0_j\rangle\}$, where matrix $\mathcal{U} \equiv \{
U_{jk}\}$ is  defined by 
\begin{equation}
|\tilde{\chi}^0_k \rangle  = \sum_{j} U_{jk}
|\tilde{\chi}^1_j\rangle,
\label{eq:cheat}
\end{equation}
i.e., the unitary linking the two ensembles \cite{HJW93}.  In summary,
perfect  concealment against  Bob allows  Alice to  cheat by  remotely
steering Bob's state to either ensemble.  More generally, relaxing Eq.
(\ref{eq:conceal}), we  may let $\rho_0^B\simeq\rho_1^B$,  which would
correspondingly diminish Alice's ability to cheat.

Interestingly, an  analogous attack on  a toy bit  commitment protocol
using      steering     correlations      can     be      demonstrated
\cite{disilvestro2017quantum}  in  the  toy  theory  due  to  Spekkens
\cite{Spe07}, which features steering but not nonlocality.

In response to  the steering attack, various works  have studied cheat
sensitive QBC, where Bob's security  requirement is relaxed by letting
$\rho_0^B\ne\rho^B_1$ (cf.  \cite{He15}, and references therein).  But
others have  questioned whether this framework  of mistrustful quantum
cryptography  is  broad enough  to  truly  rule  out secure  QBC  (see
\cite{He11, He14, he2017unconditionally, Yue12, yuen2008impossibility,
  Che15,song2017quantum} and  references therein), and our  work lends
credence to this  point of view.  A key observation  in these works is
that, unlike (say) the no-cloning  or no-signaling theorems, the no-go
argument for (non-relativistic)  QBC does not seem to  invoke a simple
and  broad physical  principle,  and  thus appears  to  be  tied to  a
specific framework.

Our  point  of departure  is  to  draw  attention to  two  assumptions
implicit in the no-go theorems in the standard framework:
\begin{description}    
\item[Assumption  (a)]  That  Alice  knows  the  two  final  ensembles
  .
\item[Assumption (b)] Quantumness  of the evidence  $E$.
\end{description}
Our approach is  to explore the possibility of  securing QBC protocols
by relaxing these two assumptions.

Regarding Assumption (a), consider a QBC protocol in which Alice lacks
full knowledge  of $\{\ket{\tilde{\chi}^\alpha_j}\}$, say  because the
two  ensembles are  determined  through secret  choices  of Bob.   Her
ignorance   may  prevent   her  from   computing  the   cheat  unitary
$\mathcal{U}$, in view of Eq.   (\ref{eq:cheat}). But, it may well end
up  over-empowering  Bob.   Thus,  we want  a  protocol  that  relaxes
Assumption (a), but with safeguards to guarantee the concealingness of
evidence $E$,  essentially via  secrecy injected  by Alice.   Thus, we
expect such a  possible protocol to be  ``double-blind''.  These ideas
are discussed  in, and form the  basis of our protocol  P1 in, Section
\ref{sec:P1}.

Regarding  Assumption (b),  a QBC  protocol in  which evidence  $E$ is
fully classical is trivially protected  against the steering attack of
\cite{May97,LC97,cDP+13}.   At first  sight, the  classicality of  $E$
would appear to  reduce such a protocol to a  classical bit commitment
scheme        \cite{goldreich2009foundations},        for        which
information-theoretic   arguments   exist  prohibiting   unconditional
security. However, irrespective of the security status of QBC protocol
obtained by relaxing Assumption (b), it is clear that such a reduction
is  not  the  case,  given  that there  will  be  intermediate  stages
involving  quantum communication  and quantum  operations.  It  may be
helpful to think of the QBC protocol with a classical evidence $E$, as
one that  implements a ``classical-valued quantum  one-way function'',
i.e., an operation with a classical output, whose difficulty to invert
comes  from quantum  nonclassicality, rather  than from  computational
complexity.

Classical  bit  commitment  based  a one-way  function  $f$  works  as
follows:  Alice  computes $f(K_{\rm  priv},  K_{\rm  pub}, a)$,  where
$K_{\rm priv},  K_{\rm pub}$ and  $a$ are Alice's private  key, public
key and  commitment. She submits  $E = \{f(K_{\rm priv},  K_{\rm pub},
a), K_{\rm  pub}\}$ as her  evidence.  Later, at the  unveiling stage,
she  submits  the  message  $\{K_{\rm priv},  a\}$.  Bob  accepts  her
commitment  after  checking  that  it reproduces  $E$.  This  is  only
computationally secure, since with sufficient computational resources,
Bob could invert $E$ to derive $a$.

In the  proposed quantum  realization of the  classical-valued one-way
function $f$, the basic idea is  that this function will have the form
$g  = f(K_{\rm  pub}, K_{\rm  priv}, L_{\rm  pub}, L_{\rm  priv}, a)$,
where  $L_{\rm pub}$  and $L_{\rm  priv}$ are  the public  and private
inputs  of Bob,  and the  output  of $f$  is generally  probabilistic.
Unlike in classical  bit commitment, here all inputs of  Alice and Bob
consist  in  general  of  quantum state  preparations,  rotations  and
measurements.  The hope  is that the privacy of Bob's  input will bind
Alice, while the  privacy of Alice's input will  provide the requisite
one-wayness  to  conceal  the  commitment.   Thus,  as  with  relaxing
Assumption (a), we require  the double-blindness feature.  These ideas
are discussed  in, and form  the basis of  our protocol P2  in Section
\ref{sec:P2}.

In summary, with  both protocols P1 and P2, Alice  is unable to launch
the steering-based attack, though  for different reasons.  In protocol
P1, this happens because Alice can't determine the malicious operation
$\mathcal{U}$ required  to steer  Bob's ensemble, whereas  in protocol
P2, this  happens because no  such $\mathcal{U}$ exists.  As  it turns
out,  P1 is  still  vulnerable to  a  less devastating,  probabilistic
attack by  Alice, based on her  superposing her commit actions.   As a
result,  protocol   P1  lacks   certification  of   classicality  (CC)
\cite{kent2000impossibility}.  On the other hand, P2 is secure even in
this stronger (CC) scenario, and is our main proposal.
  
With  a foundational  objective in  mind,  we propose  protocol P3  in
Section \ref{sec:P3}, which  is an extension of P2,  together with the
use  of  entanglement.   The  security  of protocol  P3  is  shown  to
demonstrate the reality  of the quantum state, under  the exclusion of
retrocausality.

\section{Protocol P1, and the steering-based attack\label{sec:P1}} 

A BC protocol  has three phases: a  commit phase, at the  end of which
Alice submits an  evidence of her having committed to  a specfic value
$a$;  a holding  phase,  during which  her  commitment remains  valid;
finally, an unveil phase, where  she opens her commitment, and reveals
supporting information.

The commit and unveil phases of protocol P1 are as follows.  A holding
phase of arbitrary duration between these two phases is, in principle,
allowed.  Although protocol P1 requires quantum memory for the holding
phase,  yet with  a slight  modification,  this can  be avoided.   The
protocol makes  use of quantum  encryption, which is the  task whereby
$n$ qubits can be maximally mixed  using $2n$ bits of a random private
key \cite{MTW00}.  This relies on the  fact that given any qubit state
$\rho$,  $\frac{1}{4}(\rho  +  X\rho  X  +   Y\rho  Y  +  Z\rho  Z)  =
\mathbb{I}/2$, where $X, Y$ and $Z$ are the Pauli operators.  Here, we
use  the   notation  wherein  $|0\rangle$  and   $|1\rangle$  are  the
eigenstates  of  the  Pauli  $Z$  operator,  while  $|\pm\rangle$  are
eigenstates of Pauli $X$.

\texttt{Commit  phase:}  (C1)$_{P1}$  Bob   transmits  to  Alice  $2n$
``single-blind''  (i.e.,   unknown  to  Alice)  random   qubit  states
$|\phi_j^{(\alpha)}\rangle \in \{|0\rangle, |1\rangle, |\pm\rangle\}$,
where $\alpha\in\{0,1\}$,  $0\le j  < n$, indicating  the two  sets to
her. (Alternatively, he could submit  halves of Bell states, deferring
measurement in $X$  or $Z$ basis on the ``home''  qubits until later.)
Additionally, he  supplies $Q$  extra qubits  prepared in  pure states
unknown  to  her  (or  as  halves  of  singlets),  where  $Q  \gg  n$.
(C2)$_{P1}$ Alice prepares $Q$  ``decoy'' qubits by quantum encrypting
the states of the extra qubits.  We denote the $2Q$ bits of encryption
information $\mathcal{R}_Q$.   To commit to  bit $a$, she  inserts the
$n$  states  $|\phi_j^{(a)}\rangle$ at  positions  $W$  among the  $Q$
decoys,  and  then  rearranges  all  $n+Q$  qubits  using  permutation
$\mathcal{P}$.   (C3)$_{P1}$ She  transmits  back to  Bob these  $n+Q$
qubits as evidence $E$ of her commitment.

\texttt{Unveil phase:} (U1)$_{P1}$ Alice announces $a$, $\mathcal{P}$,
$\mathcal{R}_Q$ and positions  $W$. She returns the $n$  qubits of her
non-commit  state  $\ket{\phi_j^{(\overline{a})}}$.   (U2)$_{P1}$  Bob
extracts the $n$ commit qubits from evidence $E$ using information 
$\mathcal{P}$  and  $W$.   He  verifies   that  they  are  the  states
$|\phi_j^{(a)}\rangle$.  Further, he verifies  that the $n$ non-commit
qubits    returned    in    step   (U1)$_{P1}$    are    the    states
$\ket{\phi_j^{(\overline{a})}}$.     Finally,     using    information
$\mathcal{P}$ and $\mathcal{R}_Q$,  he checks that the  $Q$ decoys are
indeed the extra qubits sent by him.  \blu \hfill $\blacksquare$ \bla

The protocol assumes the availability of quantum memory. However, this
is  not  essential. As  in  the  BB84  protocol \cite{BB84},  Bob  can
measure, just after the commit phase  ends, each evidence qubit $j$ in
a  BB84 basis  ($X$  or $Z$).   When Alice  opens  her commitment,  he
verifies  that there  is outcome  agreement  on all  qubits where  her
measurement   basis   $j$  matches   that   of   the  preparation   of
$\ket{\phi^{(a)}_j}$.   Overall,  this  doesn't affect  the  following
security  agruments, except  to roughly  halve the  security parameter
from $n$ to $\frac{n}{2}$.

Consider the security  against Bob.  Intuitively, the  large number of
Alice's  decoy qubits  swamp  the relatively  small  number of  coding
qubits $\ket{\phi_j^{(a)}}$, making it impossible for him to determine
$a$.  More quantatively, until Alice's announcement of the information
$\mathcal{P}$ and $W$ in (U1)$_{P1}$, the state of the evidence is
\begin{equation}
\rho_B^a                        =                        \mathcal{C}_W
\left[\bigotimes_j \left(\ket{\phi^{(a)}_j}\bra{\phi^{(a)}_j}\right)
  \otimes \left[\frac{\mathbb{I}}{2}\right]^{\otimes Q} \right],
\label{eq:cw}
\end{equation}
where $\mathcal{C}_W$  represents the  uniform mixture over  all ${Q+n
  \choose n}$  combinations of interpolating the  $n$ scrambled qubits
$\ket{\phi^{(a)}_j}$ among the $Q$ decoy  qubits inserted by Alice. We
note that  from Bob's viewpoint, all  decoys inserted by Alice  are in
the  state $\frac{\mathbb{I}}{2}$.   Clearly, for  any given  $n$, the
state $\rho_B^0$ approaches $\rho^1_B$  closer for larger $Q$, thereby
allowing  satisfaction  of  Eq.  (\ref{eq:conceal})  to  any  required
degree.

We indicate this showing that $\rho^a_B$,  for either $a$, can be made
arbitrarily  close  to $\left(\frac{\mathbb{I}}{2}\right)^{(n+Q)}$  in
$Q$ in  terms of the fidelity  $F(Q,n)$. 
 Then,  in  view  of  Eq.
(\ref{eq:cw}), for given $n$ and sufficiently large $Q  \gg n$,
it  follows (Appendix \ref{sec:ev}) that the fidelity  $F(Q,n) 
\equiv \mathcal{F}\left( \rho^a_B,
\left(\frac{\mathbb{I}}{2}\right)^{\otimes (Q+n)}\right)$ satisfies
\begin{align}
F(Q,n) \ge 1 - 2^{-Q(1-H(n/Q))},
\label{eq:FnQ}
\end{align}
where   $H(x)=-x\log(x)-(1-x)\log)1-x)$   is    the   Shannon   binary
entropy. For a  given $n$, fidelity $F(Q,n)$ is thus  seen to approach
unity exponentially fast as $Q$ increases, meaning that Bob can hardly
find  out  $a$.   Even   though  Bob  knows  $\ket{\phi^{(0)}_j}$  and
$\ket{\phi^{(1)}_j}$ for each $j$,  Alice's evidence for either commit
bit is close to $\mathbb{I}/2$.  There is no advantage for Bob even if
he   transmits    halves   of   Bell   state    $\ket{\Phi^+}   \equiv
\frac{1}{\sqrt{2}}(\ket{00}+\ket{11})$ since  under Alice's  method of
concealing  her commitment  by inserting  decoys is  symmetric to  any
choice of states he makes.

As  regards security  against Alice,  crucial to  protection from  the
steering attack  is the fact that  both ensembles $\mathcal{E}^\alpha$
are unknown  to Alice because  of the single-blindness  feature, i.e.,
the  dropping of  assumption (a).  But,  as we  noted earlier,  merely
Alice's  ignorance  of  the  ensembles  $\mathcal{E}^\alpha$  may  not
guarantee that she can't determine $\mathcal{U}$.

To see this, the situation can  be formalized as follows: Bob prepares
and sends  to Alice an unknown-to-her  (``single-blind'') state $\xi$.
Alice  then returns  to  Bob  an element  of  the ensemble  $T_a(\xi)$
($\alpha  \in  \{0,1\}$)  as  evidence.    As  Bob  varies  the  blind
information $\xi$, clearly the  two ensembles $T_\alpha(\xi)$ are also
rotated in some  way.  In general, $T_0(\xi)$ and  $T_1(\xi)$ need not
be indistinguishable under arbitrary variation of $\xi$.

However,       with      the       indistinguishability      condition
Eq. (\ref{eq:conceal}), the ensembles  corresponding to $T_0(\xi)$ and
$T_1(\xi)$ indeed  co-rotate in such  a way that the  steering unitary
$\mathcal{U}$ linking  them is invariant.

To prove  this, we  allow for  the encoding  ensembles to  be randomly
chosen by  Bob, say through a  secret parameter $\mu$ he  holds, i.e.,
$\mathcal{E}^{\alpha|\mu}             \equiv            \{V^\alpha_\mu
|\tilde{\chi}_j^{\alpha}\rangle\}$,   for  unitaries   $V^\alpha_\mu$.
Alice has a quantum computer, which entangles the evidence system with
an auxilliary  system she holds,  in such a  way that it  realizes the
positive operator-valued  measure (POVM)  that generates  the ensemble
$\ket{\tilde{\chi}_j^\alpha}$ for any commitment $\alpha$ she chooses.

If Bob applies  his secret transformation, then  generated ensemble is
correspondingly  rotated.   Now  if  $V^0_\mu \ne  V^1_\mu$,  then  in
general, this would imply that        
\begin{equation}
\sum_j        V^0_\mu         \ket{\tilde{\chi}_j^0}
\bra{\tilde{\chi}_j^0}V^{0\dag}_\mu      \ne       \sum_j      V^1_\mu
\ket{\tilde{\chi}_j^1} \bra{\tilde{\chi}_j^1}  V^{1\dag}_\mu,
\end{equation} 
even    if    $\rho^0_B     \equiv    \sum_j    \ket{\tilde{\chi}_j^0}
\bra{\tilde{\chi}_j^0}      =       \sum_j      \ket{\tilde{\chi}_j^1}
\bra{\tilde{\chi}_j^1}       \equiv       \rho^1_B$,      as       per
Eq. (\ref{eq:conceal}).   Therefore, in general, to  order to preserve
indistinguishablity Eq.   (\ref{eq:conceal}) when $\mu$ is  varied, we
require that the two ensembles be rotated identitcally, i.e.,
\begin{equation}
V^0_\mu = V^1_\mu \equiv V_\mu.
\label{eq:cheung}
\end{equation}
Bob's  random  choice   of  $\mu$  could  be   represented  through  a
purification, whereby we replace  $|\Psi\rangle$ in Eq. (\ref{eq:mlc})
by  $\sum_\mu \sqrt{q_\mu}\ket{b_\mu}|\Psi_\mu\rangle  \equiv \sum_\mu
\sqrt{q_\mu}\ket{b_\mu}V_\mu\ket{\Psi}       \equiv       \sum_{j,\mu}
\sqrt{q_\mu}         \ket{b_\mu}        |\phi_j^0\rangle         V_\mu
|\tilde{\chi}_j^0\rangle$, and Bob holds  the first register, which he
measures in the $\{\ket{b_\mu}\}$ basis  earlier or later.  It follows
from this  and Eq.  (\ref{eq:cheat})  that each $U_{jk}$  is unchanged
and therefore that the same cheat operator $\mathcal{U}$ works for any
$V_\mu$.   This is  so even  if  Bob measures  his entangled  register
later.   Therefore, a  straightforward extension  of the  standard QBC
framework  to allow  for ensembles  $\mathcal{E}^\alpha$ to  depend on
Bob's private  choices, doesn't automatically protect  QBC against the
steering attack (cf.  \cite{Che05, Che06}).

But in the present case,  the states chosen for $|\phi^{(0)}_j\rangle$
and those chosen for  $\ket{\phi^{(1)}_j}$ are independent and random,
with    Alice's   insertion    of    decoy    qubits   ensuring    Eq.
(\ref{eq:conceal}) to an arbitrary  degree, with $\rho^\alpha_B \simeq
\mathbb{I}/2$   for  both   values   of   $\alpha$.   Therefore,   Eq.
(\ref{eq:cheung})  is  not a  necessary  condition.   Each random  and
independent  choice of  $|\Phi^{(0)}\rangle$ and  $|\Phi^{(1)}\rangle$
would,  in  general,  determine   a  different  $\mathcal{U}$  in  Eq.
(\ref{eq:cheat}), which  Alice doesn't  know, and  hence is  unable to
deploy.

Finally, we note  that protocol P1 is secure  against simpler attacks.
Suppose   Alice   tries  the   trivial   attack   of  inserting   both
$\ket{\phi^{(0)}_j}$  and $\ket{\phi^{(1)}_k}$  among  the decoys  and
tries to  unveil what she pleases.  She will indeed be  able to unveil
the  commit state  $\ket{\phi^{(a)}_j}$  correctly, but  be unable  to
produce  the required  non-commit state  $\ket{\phi^{\overline{a}}_k}$
having already included it among the  decoys, and also be unable (with
probability  exponentially close  to 1  in $n$)  to convince  Bob that
these  purported decoys  are prepared  using the  extra qubits  he had
sent.

The requirement that  her decoys should be created  by randomizing his
extra qubits also helps thwart a port-based teleportation (PBT) attack
by  her.   Recall  that  PBT  is  a  task  wherein  she  can  teleport
(asymptotically) deterministically a qubit to port $k$ randomly picked
out   of    many   entanglement    output   ports   on    Bob's   side
\cite{ishizaka2008asymptotic, ishizaka2009quantum}. Alice might try to
send Bob  halves of singlets  instead of  the honest decoys,  and then
implement PBT  on either $\ket{\phi^{(0)}_j}$  or $\ket{\phi^{(1)}_k}$
during  the  unveiling  phase.  While  she  can  indeed  achieve  this
asymptotically, she will fail (with probability exponentially close to
1 in $Q$) to convince Bob that the states of qubits in the other ports
than $k$ were indeed prepared from the extra qubits he had supplied.

Lastly, suppose she performs the honest commitment operation for $a=0$
and naively tries to unveil $a=1$, or vice versa, then the probability
that she can escape his check  in step (U2)$_{P1}$ in general vanishes
or  is exponentially  small  in $n$,  since the  fidelity  of the  two
possible  encoding states  is $\Pi_{j=0}^{n-1}  |\langle \phi^{(0)}_j|
\phi^{(1)}_j \rangle|^2$.

Interestingly,  despite  being  impervious  to  the  steering  attack,
protocol  P1 does  not offer  a ``certificate  of classicality''  (CC)
\cite{kent2000impossibility}, i.e., the  probabilities $p_a$ to unveil
$a$     only    satisfy     the    weaker     condition    $p_0+p_1=1$
\cite{kent2000impossibility},  rather  than the  stronger  requirement
that precisely one of $p_0$ and $p_1$ should be 1 and other 0.

The reason  is that Alice would  be able to launch  an attack, wherein
she creates  a superposition of  the two commit actions  by entangling
the  quantum  computer  performing  them,  with  a  suitably  prepared
auxiliary.  We  shall refer to  this as the  ``superposition attack''.
Although less devastating than the steering attack, clearly it is more
persistent.

In this attack on protocol  P1, Alice prepares an auxiliary $A^\prime$
in  the  state   $\frac{1}{\sqrt{2}}  (\gamma_0|0\rangle_{A^\prime}  +
\gamma_1|1\rangle_{A^\prime})$, and  through a joint  interaction with
all $2n$ qubits received from Bob plus $A^\prime$, produces the state:
\begin{align}
    \sum_{a=0}^1 \gamma_a |a\rangle_{A^\prime} \otimes
         |\Phi^{(\overline{a})}\rangle_{\rm keep}
\otimes 
               |\Phi^{(a)}\rangle_{\rm encrypt},
\end{align}
where  $\sum_a   |\gamma_a|^2=1$  and  the  subscripts   ``keep''  and
``encrypt''    refer    to    the   action    of    retaining    state
$\ket{\Phi^{(\overline{a})}}$    and     transmitting    the    system
corresponding  to   $\ket{\Phi^{(a)}}$  after  insertion   of  decoys,
respectively.    Alice  measures   register  $A^\prime$   just  before
unveiling, leading  to either  $a=0$ or  $a=1$ being  chosen randomly,
with probability $|\gamma_a|^2$.

This lack of CC of the commit  bit is a feature shared with some other
proposed  QBC  protocols,  such   as  the  relativistic  BC  protocols
\cite{Ken99, Ken12, KTH+0, LKB+0, LKB+15, VMH+16} and also \cite{He11,
  He14}.  In  practice, it is  questionable whether a  dishonest Alice
would  opt  for  such  a   random  cheat  strategy  in  a  stand-alone
application of bit commitment, but  the fact remains that the protocol
doesn't  bind her  to commit  deterministically, which  undermines the
composability of this QBC protocol  in a larger application, which may
require the commitment to have a specific classical value.

We  next present  protocol P2,  that  closes the  above security  gap,
essentially   by  relaxing   Assumption  (b)   mentioned  in   Section
\ref{sec:intro}  as   being  implicit   in  the   standard  framework.
Obviously, neither  the steering  attack nor the  superposition attack
would  be  possible  if  the two  system  $E$  is  \textit{classical},
typically  encoding  classical  information  based on  outcomes  of  a
measurement determined by her commitment.  We stress that the evidence
$E$, although by itself classical,  is generated by quantum operations
(which   are    restricted   by   nonclassical   features    such   as
non-commutation, no-cloning,  measurement disturbance, etc.),  so that
the  present protocol  won't be  reducible  to a  purely classical  BC
(which  is known,  by  classical information  theoretic arguments,  to
guarantee   only  computational,   and  not   unconditional,  security
\cite{goldreich2009foundations}.)

But this would seem to endanger the concealment against Bob.  To fight
this  threat,  Alice  must  initially prepare  the  states.   However,
\textit{this}  would  introduce  the  new  threat  of  Alice's  taking
advantage of her preparation knowledge.  Therefore, Bob must randomize
these  states  in  some  way.   Thus the  encoding  states  should  be
``double-blind'',  unknown  to both  Alice  and  Bob.  These  are  the
considerations behind the following protocol.

\section{Protocol P2, and the superposition-based attack \label{sec:P2}} 

The scheme P2 can be considered (as noted in Section \ref{sec:intro}))
as a classical-valued quantum realization of a (probabilistic) one-way
function.  The  commit and  unveiling phases are  as follows,  with an
intervening holding phase of arbitrary duration. Even from a practical
standpoint,  there  is  no  time  constraint on  the  storage  of  the
evidence, since it is classical.

\texttt{Commit phase:} (C1)$_{P2}$ Alice  transmits to Bob $2n$ qubits
randomly   prepared  in   states   $|\psi_k\rangle  \in   \{|0\rangle,
|1\rangle, \ket{\pm}\}$.   (C2)$_{P2}$ Bob randomizes their  bases (by
randomly  applying either  identity $I$  or  Hadamard $H$  to each  of
them), randomizes their  bits by quantum encryption,  and finally also
randomly scrambles the qubits  according to some permutation operation
$\mathcal{P}$.  (C3)$_{P2}$ Bob transmits  to Alice these double-blind
states, denoted $|\phi_j\rangle$.  (C4)$_{P2}$  Alice picks out $n$ of
the  transmitted  states,  and  asks Bob  to  reveal  his  randomizing
operations  for these  qubits.  Upon  receiving this  information, she
verifies  that they  are indeed  qubits she  had prepared.   These $n$
check  qubits  are discarded.   (C5)$_{P2}$  To  commit to  bit  $a=0$
(resp., $a=1$), she measures the remaining $n$ states $|\phi_j\rangle$
in the basis  $Z$ (resp., $X$).  The $n$-bit random  outcome string is
denoted  $M$.   (C6)$_{P2}$  She  announces $M$  as  evidence  of  her
commitment.

\texttt{Unveil  phase:}  (U1)$_{P2}$  Alice   announces  $a$  and  her
preparation  information of  the qubits  $|\psi_k\rangle$. (U2)$_{P2}$
From the latter, Bob obtains  complete classical knowledge of all $2n$
states $|\phi_j\rangle$.  (U3)$_{P2}$ Bob verifies that the string $M$
is compatible with  the measurement of states  $|\phi_j\rangle$ in the
basis $Z$ (resp., $X$) if $a=0$ (resp., $a=1$).

Consider the security against Bob.  Prior  to the unveil phase, he has
no classical  information about  the preparation  of $|\psi_j\rangle$,
and hence  of $|\phi_k\rangle$,  which is derived  from $\ket{\psi_j}$
via  random rotations  and rearrangements,  but without  measurements.
Therefore, knowledge  of the string  $M$ reveals to him  nothing about
$a$.  Let  $\mathcal{R}$ denote the classical  information about Bob's
randomization  operations  in  step (C2)$_{P2}$  and  $H(\cdot|\cdot)$
classical conditional entropy. Then P2 satisfies the condition:
\begin{equation}
H(\alpha|M,\mathcal{R})=1,
\label{eq:hide}
\end{equation}
where $\alpha$ is the commitment random variable.  Eq. (\ref{eq:hide})
replaces Eq.  (\ref{eq:conceal}) as  the condition of security against
Bob appropriate to this scenario.  Bob can't substitute his own states
in step  (C3)$_{P2}$, nor measure  Alice's qubits, since  such actions
would generate disturbance that would  almost certainly be detected in
check (C4)$_{P2}$, where Alice verifies  that Bob has returned her own
qubits after a unitary and rearrangement operation.

As to  security against Alice, she  can't launch a steering  attack or
even a  superposition attack  for the trivial  reason that  her commit
evidence  is  now  \textit{classical}  information,  and  thus  is  in
principle  unsteerable  and  un-superimposable.   In  other  words,  a
malicious steering operation $\mathcal{U}$  simply doesn't exist. This
stands  in contrast  to  the  situation with  protocol  P1, where  the
operation $\mathcal{U}$ does exist, but Alice can't ascertain it.

Of course,  other forms  of attacks  must be  considered. Intuitively,
security  against  Alice comes  from  the  fact  that because  of  her
ignorance of $\mathcal{R}$, she is maximally ignorant of the states by
measuring which she generates string  $M$.  Thus she can't confidently
unveil a  fake measurement  basis.  Now,  even without  the scrambling
action,  the basis  and bit  randomizations ensure  that Alice  has no
information about  the returned state $\ket{\phi_j}$.   However, given
her initial preparation in step  (C1)$_{P2}$ and her final measurement
in  step (C5)$_{P2}$,  Alice  can launch  a  local entanglement  based
attack of the following kind.

In step (C1)$_{P2}$ she sends to  Bob half a singlet $|\Phi^+\rangle =
\frac{1}{\sqrt{2}}(\ket{00} +  \ket{11})$.  Her singlet will  now have
been modified through Bob's actions according to
\begin{equation}
\begin{array}{l|l}
\textrm{Bob's~action} & \textrm{Alice's~state}\\
\hline
II, IZ & \ket{00} \pm \ket{11} \\
IX, IY & \ket{01} \pm \ket{10} \\
HI, HX & \ket{0+} \pm \ket{1-} \\
HZ, HY & \ket{0-} \pm \ket{1+},
\end{array}
\label{eq:Bobstrategies}
\end{equation}
where  the  normalization  factor  has  been  dropped  out.   In  step
(C5)$_{P2}$, Alice measures in the standard Bell basis $\{\ket{00} \pm
\ket{11},  \ket{01}  \pm  \ket{10}\}$.    Suppose  she  finds  outcome
$\ket{00} -  \ket{11}$.  From Eq. (\ref{eq:Bobstrategies})  it follows
that Bob could not have applied the three operations: $II, IX, IY$.

Now  the  four honest  states  that  would  result under  Bob's  eight
possible randomization actions are:
\begin{equation}
\begin{array}{l|l|l|l|l}
\textrm{Bob} & \ket{0} & \ket{1} & \ket{+} & \ket{-} \\
\hline
II           & \ket{0} & \ket{1} & \ket{+} & \ket{-} \\
IX           & \ket{1} & \ket{0} & \ket{+} & \ket{-} \\
IY           & \ket{1} & \ket{0} & \ket{-} & \ket{+} \\
IZ           & \ket{0} & \ket{1} & \ket{-} & \ket{+} \\
HI           & \ket{+} & \ket{-} & \ket{0} & \ket{1} \\
HX           & \ket{+} & \ket{-} & \ket{1} & \ket{0} \\
HY           & \ket{-} & \ket{+} & \ket{1} & \ket{0} \\
HZ           & \ket{-} & \ket{+} & \ket{0} & \ket{1}\\
\hline
\end{array}
\label{eq:honestBob}
\end{equation}
Note that  only the  last five rows  in Eq.   (\ref{eq:honestBob}) are
applicable in  this case. Alice can  announce an arbitrary bit  as her
measurement outcome in step (C6)$_{P2}$,  say $M_j=0$. Then, to unveil
$a=0$ (resp.,  $a=1$), she  must claim  $\ket{\psi_j}=\ket{0}$ (resp.,
$\ket{\psi_j}=\ket{-}$),  since  this  would be  consistent  with  her
having applied any of these  five operations, whereas if $M_j=1$, then
to unveil $a=0$ (resp.,  $a=1$), she must claim $\ket{\psi_j}=\ket{1}$
(resp., $\ket{\psi_j}=\ket{+}$), in  view of Eq. (\ref{eq:honestBob}).
Therefore, Bob's scrambiling  action is necessary, in  addition to his
bit and basis randomizing actions.

Given the  exponentially large number  of ways to permute  the qubits,
with Bob's bit, basis and  position randomization in step (C2)$_{P2}$,
Alice  is fully  uncertain about  which of  the four  states $\ket{0},
\ket{1},  \ket{\pm}$  is each  qubit  $\ket{\phi_j}$,  even given  her
preparation  information  of   $\ket{\psi_j}$.   Therefore,  Alice  is
constrained to execute the measurements in (C5)$_{P2}$ as per protocol
and to announce  a honest $M$ at step (C6)$_{P2}$.   To cheat, at best
she  simply unveils  the  wrong  basis. This  would  be detected  with
probability  $\frac{1}{4}$  in   view  of  Eq.   (\ref{eq:honestBob}),
implying   that  her   probability   to  escape   detection  will   be
$\left(\frac{3}{4}\right)^n$.

That Alice is  forced to measure the qubit  states $|\phi_j\rangle$ in
step (C5)$_{P2}$ guarantees that protocol  P2 carries a CC. Therefore,
it can be safely composed with other instances of protocol P2 or other
tasks in a  larger application. Protocol P2 is  our principal proposal
for an experimental implementation,  though in practice the simplified
(without quantum memory) protocol P1 may suffice.

\section{Protocol P3, and the reality of the quantum state
\label{sec:P3}}

From a  foundational (rather than cryptographic)  perspective, we find
it  instructive to  consider the  following  protocol P3,  which is  a
simple extension of protocol P2. We  will show that the security proof
of P3  reduces to that  of P2,  in the sense  that the security  of P2
guarantees that of P3.

In the commit phase, the steps (C1)$_{P3}$ through (C6)$_{P3}$ are the
same  as  the  steps (C1)$_{P2}$  through  (C6)$_{P2}$,  respectively,
except:  in  (C3)$_{P3}$,  Bob  additionally  transmits  $\frac{n}{2}$
qubits, which  are halves  of singlets, to  Alice; in  (C5)$_{P3}$, in
addition to all actions of (C5)$_{P2}$ additionally Alice measures the
singlet-halves sent by Bob in the basis $Z$ or $X$ basis, depending on
whether each bit  in the second half  of $M$ is 0  or 1, respectively.
The  outcomes  for  these measurements  constitutes  $\frac{n}{2}$-bit
string  $M_2$;   and  in   (C6)$_{P3}$,  Alice  transmits   the  first
$\frac{n}{2}$ bits  of $M$  as before (denote  this string  by $M_1$).
Further,  she transmits  the $\frac{n}{2}$  bits $M_2$.  Thus, Alice's
classical  evidence $E$  in this  is the  set of  two strings  $\{M_1,
M_2\}$.

Similar, the  unveiling phase  of protocol  P3, the  steps (U1)$_{P3}$
through  (U3)$_{P3}$ are  the same  as the  steps (U1)$_{P2}$  through
(U3)$_{P2}$, respectively, except:  in (U1)$_{P3}$, Alice additionally
transmits the remaining $\frac{n}{2}$ bits of $M$; in (U3)$_{P3}$, Bob
checks  that the  outcomes  $M_2$ are  compatible  with measuring  his
halves of the singlets in the  bases specified by these remaining bits
of $M$.

Instead of a detailed proof of security of P3, it will suffice to show
that its security proof reduces to that  of P2. In protocol P3, set $n
\rightarrow  2n$, and  suppose that  Alice  and Bob  ignore the  $M_2$
part. Then,  it is clear  that the resulting  protocol is at  least as
secure as protocol P2 on $n$ bits.  If P2 is secure (as we saw it is),
then so  is P3.   We wish  to draw  attention to  the fact  that $M_2$
reaches Bob at the end of the commit phase, entailing that the singlet
halves on his side are already ``collapsed''.

We now  remark on the physical  interpretation of the security  of P3,
which is  the main  motivation behind its  proposal.  In  the standard
Hilbert space formalism, when distant  players Alice and Bob, mutually
at rest,  make measurements on  an entangled quantum state,  such that
Bob measures after  Alice in their common reference  frame, both agree
that there is a spacelike update  to the description of Bob's state as
a result of her measurement.   However, Bob's reduced density operator
remains unchanged.   This situation lies  at the heart of  the dilemma
regarding whether Bob's state's changed  status as a result of Alice's
measurement is  an \textit{objective} transformation (i.e.,  a genuine
ontological change  in the state  of Nature) or  a \textit{subjective}
transformation (i.e., just a Bayesian  update of her knowledge) of his
state.

Intriguingly,  this familiar  dilemma  becomes less  ambiguous in  the
entanglement scenario  of P3.   Let Alice  and Bob  be separated  by a
finite distance, and  at rest in each other's  reference frames.  They
have   synchronized   their  clocks,   which   keep   the  same   time
$t_A=t_B\equiv t$.   In their  common coordinate system,  the absolute
past   is   defined   by   the   event   set   $\mathcal{T}^-   \equiv
\{(t,\textbf{x})  :  t<0\}$,  the absolute  future  by  $\mathcal{T}^+
\equiv  \{(t,\textbf{x})   :  t>0\}$  and  the   absolute  present  by
$\mathcal{T}^0     \equiv    \{(0,\textbf{x})\}$.      Further,    let
$\mathcal{W}_B$  denote the  world line  of Bob,  and $\mathcal{W}^-_B
\equiv   \mathcal{T}^-\cap\mathcal{W}_B$,    $\mathcal{W}^+_B   \equiv
\mathcal{T}^+\cap\mathcal{W}_B$, i.e., the past and future segments of
Bob's   world   line,   and    let   event   $\mathcal{W}_B^0   \equiv
\mathcal{T}^0\cap\mathcal{W}_B$.

In  this frame,  we denote  by $\textbf{e}_A\equiv(0,\textbf{0})$  the
event  of  Alice's  entanglement  breaking (EB)  via  measurements  on
singlet halves in  step (C5)$_{P3}$.  In this instant,  she knows that
she has  ``collapsed'' --i.e., irreversibly prepared--  Bob's state in
favor of one particular commitment.  We denote this remote preparation
event in $\mathcal{W}_B$ by $\textbf{e}_B$.


Is  this  preparation  of  Bob's  state  at  event  $\textbf{e}_B$  an
objective  or subjective  transformation?   To answer  this, we  shall
employ an operational framework, i.e., one not based on an ontological
model,  but  simply  on  the   inputs  and  measured  outputs  of  the
communication  involved.  First,  we define  a feature  $X$ associated
with a system  $S$ as \textit{operational} if $X$ can  be described in
terms   of   probabilities   for    outcomes   of   quantum   physical
measurements. Features that must refer  to an ontological model for QM
aren't  operational.   A  transformation  \textbf{T}  is  said  to  be
\textit{objective} if  there is an  operational feature $X$  of system
$S$ such  that $X$  is altered  under \textbf{T}.   Ordinary classical
transformations are manifestly objective,  but this framework allows a
larger set of transformations to  be characterized as objective.  More
precisely, it enlarges the set of objective transformations to include
not just \textit{detectable} transformations, but also its superset of
\textit{verifiable}  transformations.   All classical  transformations
are detectable, and hence also  verifiable. (For example, if the color
of a classical objected turned from red to blue, this may be verified,
and indeed detected.)   But, we shall find that  quantum theory allows
verifiable transformations that aren't detectable.

We  define an  ``EPR-certificate'' as  the $2  \times \frac{n}{2}$-bit
outcome and basis  specification (namely, whether the basis  is $Z$ or
$X$) for a set of $\frac{n}{2}$-qubit pure states, that can pass Bob's
check for  outcome matching in step  (U3)$_{P3}$ \textit{with complete
  certainty}, in  support of some  commitment.  In the  terminology of
\cite{EPR35}, the EPR-certificate associates an ``element of reality''
to the  commitment encoded  in the $\frac{n}{2}$  home qubits  of Bob.
Let  $\mathfrak{p}(a,C)$  denote  the  probability  that  a  candidate
EPR-certificate  $C$  can  (upon  being  checked)  pass  the  test  in
(U3)$_{P3}$,  and further,  that  $\mathfrak{p}^\ast(a) \equiv  \max_C
\mathfrak{p}(a,C)$.   Clearly, the  probability $\mathfrak{p}^\ast(a)$
so defined is an operational feature associated with Bob's state.

Now, the security of protocol P3  implies that after Alice's EB event,
there exists  a specific EPR-certificate  for commitment $a$  and none
for $\overline{a}$.  Note that this is so even though she doesn't have
the   EPR-certificate's   complete   classical   description.    Thus,
$\mathfrak{p}^\ast   (a|\mathfrak{B})=1$   whilst   $\mathfrak{p}^\ast
(\overline{a}|\mathfrak{B}) < 1$, which entails that:
\begin{equation}
\mathfrak{p}^\ast(0|\mathfrak{B})                                  \ne
\mathfrak{p}^\ast(1|\mathfrak{B})
\label{eq:yes}
\end{equation}
where where the overline on  $\mathfrak{B}$ indicates negation,
and $\mathfrak{B}$ is the statement asserting that Alice executed an
EB in step  (C5)$_{P3}$.  (We shall ignore the effect  of noise, since
we are  concerned with  the situation  in principle.) 

When Alice hasn't yet executed this step, she knows that her choice is
uncorrelated with  Bob's preparation  in the  strong sense  that there
aren't \textit{any} EPR-certificates in the universe. Thus,
\begin{equation}
\mathfrak{p}^\ast(0|\mathfrak{B}) = \mathfrak{p}^\ast(1|\mathfrak{B}).
\label{eq:no}
\end{equation}
Bob's state  is symmetric with  respect to both  possible commitments.
This symmetry is  also consistent with the assumption  of Alice's free
will, in that her choice is unrelated to Bob's preparation.

\begin{figure}
  \begin{center}
    \begin{tikzpicture}[scale=1.25]
     \draw [thick,->] (2,-2) -- (2,4); 
     \draw [thick,->] (4,-2) to (4,4);
     \draw (2,-2) circle node [black,below] {Alice};
     \draw (4,-2) circle node [below] {Bob};
     \draw (2,4) circle node [black,left] {$t_A$};
     \draw (4,4) circle node [left] {$t_B$};
     \draw [thick,dotted,<->] (4.1,1.1) to (4.1,2.9);
     \draw [blue,fill] (4.1,-0.2) circle [radius=0] node [black,right] {$\omega_{\downarrow}$};
     \draw [thick,dotted,<->] (4.1,0.9) to (4.1,-0.9);
     \draw [blue,fill] (4.1,2.2) circle [radius=0] node [black,right] {$\omega_{\uparrow}$};
     \draw [dashed,->] (2,1) to (3.8,1);
     \draw [blue,fill] (2,1) circle [radius=0.1] node [black,left=1] {$\textbf{e}_A$};
     \draw [blue,fill] (3,1.0) circle node [black,above] {$d_\rightarrow$};
     \draw [blue,fill] (3,2.2) circle node [black,above] {$c$};
     \draw [blue,fill] (3,-0.2) circle node [black,below] {$-c$};
     \draw [thick,->] (2.1,1.1) to (3.9,2.9); 
     \draw [thick,->] (2.1,0.9) to (3.9,-0.9);  
     \draw [blue,fill] (4,3) circle [radius=0.1] node [black,right=1] {$\textbf{e}_B=\textbf{e}_B^{\uparrow}$};
     \draw [blue,fill] (4,-1) circle [radius=0.1] node [black,right=1] {$\textbf{e}_B=\textbf{e}_B^{\downarrow}$};
     \draw [blue,fill] (4,1) circle [radius=0.1] node [black,right=1] {$\textbf{e}_B=\mathcal{W}_B^{0}$};    
    \end{tikzpicture}
\end{center}
\caption{Committer Alice  and recipient  Bob are distant  observers at
  rest in each other's reference  frame. The two vertical lines denote
  their respective worldlines.  Alice's free  choice of bit $a$ at the
  EB  event  $\textbf{e}_A=(0,\textbf{0})$  prepares Bob's  system  at
  event  $\textbf{e}_B$  on his  world  line.   Possible locations  of
  $\textbf{e}_B$  include  $\mathcal{W}^0_B$   (in  Alice's  present),
  $\textbf{e}_B^\uparrow$     (in    the     causal    future)     and
  $\textbf{e}_B^\downarrow$ (in the causal past).}
\label{fig:P3}
\end{figure}
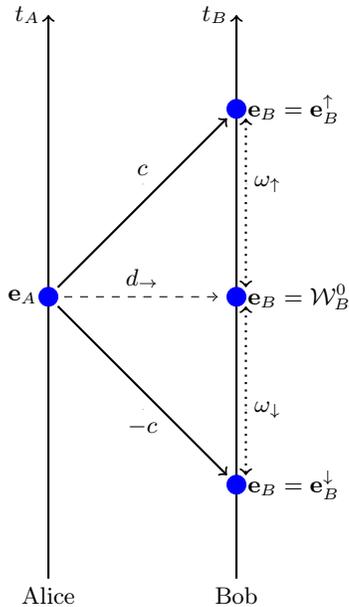

In  the  present  framework, Eqs.   (\ref{eq:yes})  and  (\ref{eq:no})
together imply that  that Alice's EB measurement  induces an objective
transformation of Bob's state, whereby  the original symmetry in Bob's
state with respect to both commitments gets verifiably broken.

The  objective nature  of the  symmetry breaking  event $\textbf{e}_B$
signifies  that  Alice's  remote  preparation  must  correspond  to  a
definite  spacetime event  on  Bob's world  line.   Where to  position
$\textbf{e}_B$ on Bob's world line?  From Alice's perspective, in view
of the constraints imposed  by Eqs.  (\ref{eq:yes}) and (\ref{eq:no}),
$\textbf{e}_B$  must be  identified  precisely with  $\mathcal{W}_B^0$
(see  Figure   \ref{fig:P3}),  which   is  spacelike   separated  from
$\textbf{e}_A$.

Observers  in relative  motion with  respect to  Alice may  make other
claims,  complicating  the picture,  to  which  we return  in  Section
\ref{sec:retro}. Now,  consider an alternative to  the above scenario,
as seen in the Alice/Bob reference frame.

\subsection{Forward-time  influence
\label{sec:forward}}  

Suppose $\textbf{e}_B$  is positioned  later on his  worldline (region
$\mathcal{W}^+_B$),   say  at   $\textbf{e}_B^{\uparrow}$,  which   is
lightlike or timelike separated from $\textbf{e}_A$.  Then, there is a
worldline segment $\omega_{\uparrow}$  between $\mathcal{W}_B^{0}$ and
$e_B^{\uparrow}$   (Figure  \ref{fig:P3})   for   which  neither   Eq.
(\ref{eq:yes})  nor Eq.   (\ref{eq:no}) would  hold true.   This would
correspond   to    a   breakdown   in   quantum    correlations   (cf.
\cite{SBB+08}), and in general lead  to violation of conservation laws
for spin angular momentum, etc.   The possibility of such breakdown is
experimentally ruled  out by  data from loophole-free  Bell inequality
violation tests \cite{HBD+15}.


Barring retrocausality  (discussed below in  Section \ref{sec:retro}),
Alice thus  concludes that  her remote preparation  of Bob's  state is
associated  with a  spacelike influence  occuring across  the interval
$d_\rightarrow$  (Figure \ref{fig:P3}).   There is  no overt  conflict
with relativistic  no-signaling, since  Bob can't  unilaterally detect
this influence  at $\textbf{e}_B$,  but can only  verify it  later on.
Worded  differently,  this  superluminal influence  corresponds  to  a
\textit{verifiable signal}, but not a \textit{detectable signal}.

We  denote  by  \textbf{VS}   (``verifiable-signaling'')  the  set  of
two-party protocols,  such as  P3, which  permit a  verifiable signal.
For protocols in  this class, one can construct a  ``causal story'' at
the operational level.  In the story associated with P3,  Alice is the
sender and Bob the receiver.

A class of  two-party protocols that is strictly weaker  is the set of
tests  of nonlocality  in quantum  mechanics or  general non-signaling
probability theories.  Here, the  correlations don't admit a ``clean''
causal story  in the  sense that  there is no  unequivocal case  for a
definite  time   ordering  behind   the  observed   correlations  (cf.
\cite{suarez2001is}).    At  best,   one   can   show  that   nonlocal
correlations observed  in these protocols,  if ``extended'' to  a more
deterministic ontological  model, would entail signaling  at the ontic
level (with no preference for Alice  or Bob in regard to the direction
of  causation)  \cite{AS16}.   These protocols  constitute  the  class
\textbf{XS}, or ``extension-signaling''.

A strictly  stronger class  than \textbf{VS} is  the set  of two-party
protocols in any of which  Bob can unilaterally detect distant Alice's
input.     These    protocols     are    denoted    \textbf{S}    (for
``detectable-signaling'',   or  simply,   ``signaling'').   Obviously,
protocols \textbf{S} are prohibited by  both QM and special relativity
when Alice's and Bob's measurements are spacelike separated.

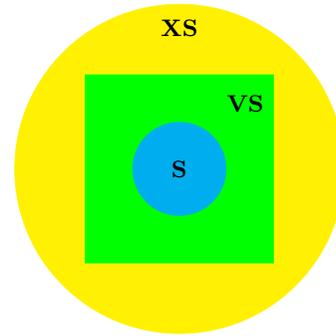
\begin{figure}
  \begin{center}
    \begin{tikzpicture}[scale=1.25]
    \draw [yellow,fill] (0,0) circle [radius=1.75];
    \draw [green,fill] (-1,-1) rectangle (1,1);
    \path [cyan,fill] (0,0) circle [radius=0.5];
    \node (0,0) {\textbf{S}};
    \node at (0.7,0.7) {\textbf{VS}};
    \node at (0,1.5) {\textbf{XS}};
    \end{tikzpicture}
\end{center}
\caption{The  containments   among  different  classes   of  two-party
  protocols,     \textbf{S}    (detectable-signaling),     \textbf{VS}
  (verifiable-signaling)   and    \textbf{XS}   (extension-signaling).
  Special   relativity  excludes   class  \textbf{S}.    Protocols  in
  \textbf{VS} admit an operational  ``causal story'', whereas those in
  $\textbf{XS}   \textbackslash  \textbf{VS}$   don't.   The   classes
  \textbf{S},  \textbf{VS}  and  \textbf{XS}   may  be  considered  as
  signaling   analogues  of   the  computational   complexity  classes
  \textbf{P}, \textbf{NP} and \textbf{PSPACE}.}
\label{fig:classes}
\end{figure}

The  containments  among  these  classes of  two-party  protocols  are
depicted in Figure  \ref{fig:classes}.  It may be helpful  to think of
\textbf{S}, \textbf{VS} and \textbf{XS}  as signaling analogues of the
computational complexity classes \textbf{P}  (the class of efficiently
solvable  decision  problems),  \textbf{NP}  (the  class  of  decision
problems  that are  efficiently verifiable)  and \textbf{PSPACE}  (the
class  of problems  solvable in  polynomial amount  of storage  space)
\cite{petting2016zoo}, respectively.

The verifiable  signal in a  \textbf{VS} protocol, since it  causes an
objective transformation of Bob's state,  requires a causal channel to
mediate it. (If the transformation  were only subjective, then clearly
there  is  no  such  requirement.)   However,  relativistic  causality
forbids any dynamic  mechanism being responsible for it.   We are thus
led to conclude that the  quantum state vector itself, whose reduction
forms the  basis for  predicting the  remote preparation,  must supply
this  causal channel.   In  this sense,  the state  vector  is a  real
entity.


Some points  are worth noting  here.  Firstly, observe that  we deduce
the  reality  of  the  quantum state  appealing  only  to  operational
criteria,   and  without   recourse   to   an  ontological   framework
\cite{leifer2014is}.  Secondly,  our argument  for the reality  of the
quantum  state didn't  require  quantum nonlocality  but only  quantum
teleportation, which  is a  weaker resource.  Indeed,  local entangled
quantum  states  can  still  teleport  above  the  classical  fidelity
threshold \cite{barrett2001implications}. Thirdly,  the essence of our
argument could be couched in  a teleportation setting without invoking
bit commitment. The  latter mainly serves to provide  a situation that
is amenable in this framework  for defining the objectivity of Alice's
remote preparation of Bob's state.

In regard to the second point above, verifiable signaling can be shown
to  occur  even  in  a  theory  without  nonlocality,  but  permitting
teleportation, such  as Spekkens'  toy theory \cite{Spe07},  where the
state  of  a  particle  can  be modelled  epistemically  (i.e.,  as  a
probability  distribution over  certain  ontic elements).   It can  be
shown that  both the  steering-based attack on  bit commitment  in the
standard  framework  \cite{disilvestro2017quantum},  as  well  as  our
solution to this problem via protocols  P1, P2 and P3, are possible in
this toy  theory.  Our  above argument  for the  reality of  the state
vector, adapted to a toy version of protocol P3, isn't contradicted by
the epistemicity  of the  toy state, but  rather implies  that Alice's
choice at $\textbf{e}_A$ produces  a remote \textit{ontic} disturbance
at $\textbf{e}_B$,  i.e., a remote  disturbance in the  ontic elements
underlying Bob's state.



\subsection{Retrocausal influence \label{sec:retro}}  

Backward-time influence has been considered as a viable alternative to
superluminal  influence because  it  offers  attractive features  like
time-symmetry,   Lorentz   invariance  and   retaining   local-realism
\cite{price2012does, werbos2016analog, leifer2016is}.   In the present
situation, note  that the forward-time  and also the  simultaneous (in
the reference  frame of  Alice/Bob) spacelike influence,  discussed in
Section \ref{sec:forward},  would lead  certain observers  in relative
motion to Alice  to expect a breakdown in quantum  correlations of the
type  discussed above  or in  \cite{SBB+08}.  These  are observers  in
whose reference frame events in the segment $\omega_\downarrow$ happen
before  $\textbf{e}_A$. To  prevent  this, one  can  posit that  these
causal influences are transmitted through a wider future cone than the
future  light-cone, but  the required  ``speed of  information'' would
have  to be  several orders  faster than  light speed  to account  for
current  experimental verification  of  quantum nonlocal  correlations
(cf. \cite{SBB+08}, and references therein).

But, if one swallows the  bitter pill of backward-time causation, then
a  covariant solution  to  avoid predicting  a  breakdown of  nonlocal
correlations  in  any reference  frame,  is  to further  broaden  this
``wider  causal  cone''  until  it coincides  with  the  \textit{past}
light-cone rooted at $\textbf{e}_A$, so  that the verifiable signal is
transmitted  (or, state  vector ``collapses'')  along the  boundary of
event  $\textbf{e}_A$'s causal  past  \cite{kraus1970formal}.  In  the
context of protocol P3, $\textbf{e}_B$ would be positioned backward on
Bob's  worldline  at  $\textbf{e}_B^{\downarrow}$  on  $\mathcal{W}_B$
which   is    lightlike   separated   from    $\textbf{e}_A$   (Figure
\ref{fig:P3}).   This  has  the  effect of  extending  the  domain  of
validity  of  Eq.   (\ref{eq:yes})  backwards to  include  the  region
$\omega_\downarrow$.    However,  the   protocol  classes   in  Figure
\ref{fig:classes}  carry over,  though with  the ``signaling''  in the
context  of  \textbf{VS}  and  \textbf{XS}  should  be  understood  as
retrocausal influences.

Such retrocausality would  infringe on the definition  of Alice's free
will, since it  would imply that Bob's ensemble in  the causal past is
correlated with Alice's  choice.  Presumably, one can  try to redefine
free  will in  this situation  to exclude  such correlations  from its
definition.

If we accept  this retrocausal model, then one  can presumably imagine
some dynamic effect  propagating at light-speed backward  in time from
$\textbf{e}_A$ and preparing Bob's state at $\textbf{e}^\downarrow_B$,
and the above  argument for attributing reality to  the quantum state,
no  longer holds.   Hence, we  require  the assumption  of absence  of
retrocausality  to arrive  at our  conclusion  of the  reality of  the
quantum state.

\section{Conclusions and discussions}

The no-go  theorem for  quantum bit commitment  (QBC) in  the standard
non-relativistic framework is a consequence of the fact that Alice can
exploit quantum steering to unveil  either commit bit on Bob's system,
if  he  can't  distinguish  the  mixtures  corresponding  to  the  two
commitments \cite{May97, LC97, cDP+13}.  However, various authors have
questioned whether this framework is general enough to cover, and thus
rule out, all possibilities for QBC \cite{He11, He14,Yue12,Che15}.  In
line with this argument, here  we identify two assumptions implicit in
the standard framework: (a) that  Alice's submitted evidence exists in
an ensemble known  to her; and (b)  that $E$ is a quantum--  and not a
classical-- system.

Relaxing  the  assumption (a),  we  construct  a QBC  protocol  (named
``P1''), whose security against  the steering-based attack arises from
the  fact that  Alice is  unable to  determine the  malicious steering
operator $\mathcal{U}$.   However, protocol  P1 still allows  Alice to
attack probabilistically, and as a result, P1 lacks ``certification of
classicality''  (CC)  \cite{kent2000impossibility},   in  common  with
various  relativistic bit  commitment  protocols \cite{Ken99,  Ken12,
  KTH+0, LKB+15, VMH+16}.

Relaxing the  second assumption  above, we  present a  second protocol
(named  ``P2'')  that  is  secure   in  the  stronger  sense  of  also
guaranteeing CC.   Because the submitted evidence  is classical, there
trivially  exists no  malicious  steering  operator $\mathcal{U}$  and
furthermore,   the   superposition-based   probabilistic   attack   is
impossible.  In both protocols, security against simpler attacks makes
use of  the single-blindness  or double-blindness feature  through the
use  of randomization  of the  state by  quantum encryption,  particle
rearrangement,  etc.  Protocol  P2 is  most suitable  for experimental
implementation, though the quantum-memory-less  variant of protocol P1
may be practically sufficient.

Finally, we  propose a  third protocol  (named ``P3''),  which 
relaxes both assumptions in a teleportation setting,
and  is  motivated for  a  foundational  purpose.  Here, the  role  of
entanglement  and  teleportation  in  protocol P3  is  vital  for  the
argument of the  reality of the quantum state.  Alice's  free will and
the  protocol's  security  are  invoked   to  argue  that  her  remote
preparation  of Bob's  system is  an objective  transformation in  the
weaker sense  that she  produces a \textit{verifiable}  preparation of
Bob's state  (leading to cryptographic  security), rather than  in the
stronger sense of being  unilaterally \textit{detectable} (which would
have  led to  superluminal  signaling).  It  is  argued that,  barring
retrocausality,  this   remote  objective  transformation   entails  a
superluminal  \textit{influence}  which,  by  virtue  of  relativistic
causality, can't be  pinned down on any dynamical  mechanism.  This is
used  as the  basis to  argue for  the objectivity  of Alice's  remote
preparation, and hence for the reality of the quantum state.

All  our protocols  are conceptually  and experimentally  simple. They
allow  for a  holding phase  of an  indefinite time  period even  with
current  technology,  in contrast  to  the  relativistic BC  protocols
\cite{Ken99,  Ken12,   KTH+0,  LKB+15,   VMH+16},  which   require  an
increasingly complex,  continued and carefully timed  communication to
extend the  holding phase  (the current  experimental record  being 24
hours).

Finally, it  is hoped that  our results  open up new  possibilities in
mistrustful  quantum  cryptography,  in particular,  highlighting  the
difficulty in-- and consequent care  needed for-- determining the full
scope of the most general framework appropriate to mistrustful quantum
cryptography. It  also uncovers  a basic relationship  between secrecy
and the  nature of physical laws,  which would be useful  for devising
cryptographic axioms to derive quantum mechanics.

\acknowledgments

The author  thanks DST-SERB,  Govt.  of  India, for  financial support
provided through the project EMR/2016/004019.

\appendix

\section{State of the evidence \label{sec:ev}}

We conservatively  assume that  all states $\ket{\phi^{(a)}_j}$  for a
given $a$ are identical, say $\ket{0}$.  Under the stated assumptions,
Alice's evidence is in the state
\begin{align}
\rho_B^a                        &=                        \mathcal{C}_W
\left[\left(\ket{0}\bra{0}\right)^{\otimes n}
  \otimes \left[\frac{\mathbb{I}}{2}\right]^{\otimes Q} \right],\nonumber\\
&= \left( 2^{Q+n} - \sum_{j=1}^{n}  {Q+n  \choose   Q+j} \right)^{-1} \mathbb{I}^\ast \nonumber \\
&= \left( 2^{Q+n} - \sum_{j=0}^{n-1} {Q+n \choose  j} \right)^{-1} \mathbb{I}^\ast \nonumber \\
\label{eq:cw0}
\end{align}
where $\mathbb{I}^\ast$  is the  density matrix  in the  Hilbert space
$\mathcal{H}_2^{\otimes (Q+n)}$ of $2^{Q+n}$ qubits, which is diagonal
and  equal-weighted in  the  computational basis,  with precisely  the
components with Hamming weight greater than $Q$ vanishing.

For  a fixed  integer $t$,  and  integer $T  \rightarrow \infty$,  the
truncated binomial series satisfies the bound \cite{lugo2017sum}:
\begin{align}
\lim_{T\rightarrow \infty} &{T \choose t}^{-1} \sum_{j=0}^t {T \choose j} = {{T \choose t} + {T \choose t-1} + {T \choose t-2}+\dots   
\over {T \choose t}} \nonumber\\
&= {1 + {t \over T-t+1} + {t(t-1) \over (T-t+1)(T-t+2)} + \cdots}
\nonumber\\
&\le {1 + {t \over T-t+1} + \left( {t \over T-t+1} \right)^2 + \cdots}
\nonumber\\
& = \frac{T-t+1}{T-2t+1}.
\end{align}
Setting  $T \equiv  n+Q$  and $t  \equiv n-1$  here,  one finds
\begin{align}
\sum_{j=0}^{n-1} {n+Q \choose  j} 
\le {n+Q \choose n-1}\frac{Q+2}{Q-n+3}.
\label{eq:bound}
\end{align}
Substrituting this in  Eq. (\ref{eq:cw0}), we find that,  for large $Q
\gg n$,  the number of  non-vanishing entries in  $\mathbb{I}^\ast$ is
bounded   below   by   $\upsilon(Q,n)  \equiv   2^{Q+n}-{n+Q   \choose
  n-1}\frac{Q+2}{Q-n+3}  \approx  2^{Q+n}-{Q+n  \choose  n-1}  \approx
2^{Q+n} -  2^{(Q+n)H(n/Q)} = 2^{Q+n}(1 -  2^{-(Q+n)[1-H(n/Q)]}$, where
the Stirling  approximation ${N \choose  Np} \approx NH(p)$,  has been
used.

The  fidelity  between   states  $\rho$  and  $\sigma$   is  given  by
Tr$(\sqrt{\sqrt{\sigma}\rho  \sqrt{\sigma}}$.  Setting  $\sigma \equiv
2^{-(Q+n)}\mathbb{I}$    and   $\rho    \equiv   \rho^a_B$    in   Eq.
(\ref{eq:cw0}) in the above approximation,  we have fidelity $F(Q,n) =
2^{-(Q+n)/2}{\rm              Tr}(\sqrt{\rho^a_B})             \gtrsim
2^{-(Q+n)/2}\sqrt{\upsilon(Q,n)}$,    from    which,    one    obtains
Eq. (\ref{eq:FnQ}).

\bibliography{qvanta}
 
\end{document}